\documentclass[conference]{IEEEtran}

\IEEEoverridecommandlockouts
\usepackage{tabulary}
\usepackage{booktabs}
\usepackage{setspace}
\usepackage{amsmath}
\usepackage{bbm}
\usepackage[dvips]{color}
\usepackage{comment}
\usepackage{todonotes}
\usepackage{epsf}
\usepackage{epsfig}
\usepackage{times}
\usepackage{epsfig}
\usepackage{graphicx}
\usepackage{bbold}
\usepackage{mathrsfs}
\usepackage{amssymb}
\usepackage{pdfpages}
\usepackage{epstopdf}
\usepackage{tcolorbox,bbold}
\usepackage{authblk}
\usepackage{algorithm,algorithmicx,algpseudocode}
\makeatletter
\newcommand{\algmargin}{\the\ALG@thistlm}
\makeatother
\algnewcommand{\parState}[1]{\State%
  \parbox[t]{\dimexpr\linewidth-\algmargin}{\strut #1\strut}}
\usepackage{url}
\usepackage{dsfont}
\usepackage{lettrine} 
\usepackage{amsmath,epsfig,amssymb,algorithm,algpseudocode,amsthm,cite,url}
\usepackage{subcaption}
\allowdisplaybreaks
\usepackage{csquotes}

\usepackage{verbatim}
\usepackage[english]{babel}
\usepackage{amsmath,amssymb}

\usepackage[justification=centering]{caption}
\usepackage{verbatim}

\begin{document}

\title{Catch Me If You Can: Deep Meta-RL for Search-and-Rescue using LoRa UAV Networks}
\author{Mehdi Naderi Soorki\IEEEauthorrefmark{1}, Hossein Aghajari\IEEEauthorrefmark{1}, Sajad Ahmadinabi\IEEEauthorrefmark{1},  Hamed Bakhtiari Babadegani\IEEEauthorrefmark{1},\\
 Christina Chaccour\IEEEauthorrefmark{2}, Walid Saad\IEEEauthorrefmark{2}\vspace*{0.05cm}\\
\IEEEauthorrefmark{1}IWiN Research laboratory, Engineering Faculty,Shahid Chamran University of Ahvaz,Ahvaz, Iran,\\
\IEEEauthorrefmark{2}Wireless@ VT, Bradly Department of Electrical and Computer Engineering, Virginia Tech, Blacksburg, VA USA,\\
{Emails:m.naderisoorki@scu.ac.ir,\{hn.aghajari,sajadahmadinabi,h.bakhtiaribabadegani\}@gmail.com,}\\
\{christinac, walids\}@vt.edu.
}
\maketitle
\thispagestyle{empty}

\begin{abstract}
Long range (LoRa) wireless networks have been widely proposed as a efficient wireless access networks for the battery-constrained Internet of Things (IoT) devices. In many practical search-and-rescue (SAR) operations, one challenging problem is finding the location of devices carried by a lost person. However, using a LoRa-based IoT network for SAR operations will have a limited coverage caused by high signal attenuation due to the terrestrial blockages especially in highly remote areas. To overcome this challenge, the use of unmanned aerial vehicles (UAVs) as a flying LoRa gateway to transfer messages from ground LoRa nodes to the ground rescue station can be a promising solution. In this paper, an artificial intelligence-empowered SAR operation framework using UAV-assisted LoRa network for different unknown search environments is designed and implemented. The problem of the flying LoRa (FL) gateway control in the search-and-rescue system using the UAV-assisted LoRa network is modeled as a partially observable Markov decision process. Then, a deep meta-RL-based policy is proposed to control the FL gateway trajectory during SAR operation. For initialization of proposed deep meta-RL-based policy, first, a deep RL-based policy is designed to determine the adaptive FL gateway trajectory in a fixed search environment including a fixed radio geometry. Then, as a general solution, a deep meta-RL framework is used for SAR in any new and unknown environments to integrate the prior FL gateway experience with information collected from the other search environments and rapidly adapt the SAR policy model for SAR operation in a new environment. The proposed UAV-assisted LoRa network is then experimentally designed and implemented. To analyze the performance of proposed framework in real world scenarios, the proposed SAR system is tested in two different target areas: a wide plain and a slotted canyon at Mongasht mountain ranges, Iran. Practical evaluation results show that if the deep meta-RL-based control policy is applied instead of the deep RL-based one, the number of SAR time slots decreases from 141 to 50. Moreover, the average distance between UAV trajectories under deep meta-RL and deep RL based policies from the UAV trajectory under optimal policy are respectively $619$ and $1930$ meter during the SAR operation time.
\end{abstract}
\begin{IEEEkeywords}
LoRa technology, Unmanned aerial vehicle, Deep meta-reinforcement learning, Search-and-rescue operation
\end{IEEEkeywords}
\section{Introduction}\label{sec:Intro}
Unmanned aerial vehicles (UAVs) are playing an increasingly important role in next-generation wireless networks such as 5G and beyond \cite{my6Gpaper}. For instance, UAVs can  guarantee high-speed and ultra-reliable connectivity while also extending the cellular network coverage to three-dimensional (3D) space~\cite{int1}. In particular, we can temporarily move UAVs to cover Internet-of-Thing (IoT) devices and establish communications therein without high-cost conventional network infrastructures. In this regard, UAV-assisted wireless networks can decrease the operational expenditures and improve the efficiency of various IoT applications.

With the proliferation of UAV-assisted wireless access networks, our reliance on IoT applications such as smart farming, smart factory, and public safety will be more pronounced~\cite{8879484}. However, to support this IoT trend, a reliable wireless access technology with wide reach and low power consumption is required. In this regard, the so-called long-range (LoRa) communication protocol has been proposed as a promising technology for high energy-efficient and long-range communication~\cite{8879484,L_Alliance}.  These two characteristics make LoRa technology an appropriate solution for battery-constrained IoT devices that are often deployed in dispersed rural areas. A typical LoRa-based IoT network begins with a LoRa-enabled embedded sensor node that sends data to the LoRa gateway. Then, data can be sent from LoRa gateway over cellular network and then routed to application servers located at the network core. One of the key challenges of LoRa-based IoT networks is localization for outdoor environments that is needed for different applications such as navigation and tracking, air traffic control, remote sensing, intelligence, surveillance, and reconnaissance, and search-and-rescue (SAR) operations~\cite{8827665}.

Existing localization techniques are mainly based on the time difference of arrival (TDOA)  and the received signal strength index (RSSI) schemes in wireless LoRa networks~\cite{8568252}. In the so-called TDOA portioning methods with LoRa networks, the distances between a LoRa node and each LoRa gateway are estimated through a time of arrival in trilateration approach~\cite{8986229,8568252}. Thus, this method requires the use of a precise clock to synchronize between all LoRa nodes~\cite{8923542}. This implies additional communication overheads and higher and, thus, this solution is not appropriate for low-power and low-cost LoRa device~\cite{8923542}. In the RSSI trilateration positioning methods, the end-device location is estimated by RSSI value when it transmits data to the LoRa gateways without requirement of clock synchronization. Thus, RSSI-based techniques are employed to develop positioning functions using RSSI in LoRa networks~\cite{8986229}.

Several recent works such as in~\cite{8827665} and~\cite{8115843,9798060,9591156,8923542} analyze RSSI-based LoRa localization system for different scenarios. In~\cite{8827665}, the authors proposed six new RSSI-based localization algorithms to reduce the effect of non-Gaussian noise in LoRa networks by either eliminating bad anchor nodes or selecting good anchor nodes during localization. In this work, the performance of all localization algorithms is investigated using simulation model with real-data measurement with developed LoRa localization system. In~\cite{8115843}, since noise-like electronic interference and blocking can affect the accuracy of localization, the authors propose a new approach to improve the performance of a LoRa-based localization system in noisy outdoor environments. Specifically, the work in~\cite{8115843} developed two new localization algorithms based on a traditional localization linear model. The first new localization algorithm locates the noisy measurement using k-mean clustering and then re-calculates the localization outcomes without a node thereby deriving the largest estimated RSSI error. The second algorithm in~\cite{8115843} requires the localization error is low if the estimated RSSI errors of the estimated location of the target node to other anchor nodes are small. Then, the best solution is chosen by calculating the estimated RSSI errors in all possible estimated locations. In~\cite{9798060}, the authors combine fingerprint-based and model-based RSSI methods to solve the outdoor positioning problem. They adopt an interpolation-based approach to build a 3D model with 36 RSSI sampling points to achieve localization model with higher accuracy. The work in~\cite{9591156} proposed an RSSI-based method to accurately identify the location of a vehicle, equipped with a LoRa node, travelling along a known path which is divided into segments of length equal to or shorter than the desired accuracy. Values of the RSSI measured by the LoRa gateways are collected and used to characterize each segments. In~\cite{8923542}, the RSSI-based method is proposed for the localization of cattle collars communicating with LoRa radios. In particular, the authors developed an RSSI-based distance estimation using realtime adjustment of RSSI-distance mapping, taking advantage of communication between collar nodes and gateway. However, the works in~\cite{8827665} and~\cite{8923542,8115843,9798060,9591156} are based on the RSSI method and, thus, they need to deploy large number of anchor points on a large scale outdoors for SAR operation which in not practical in highly remote areas. Moreover, the works in~\cite{8827665} and~\cite{8923542,8115843,9798060,9591156} do not investigate the potential of a UAV-assisted LoRa networks.

Fortunately, employing UAV as a flying gateway in localization and tracking system can bring many attractive advantages due to its high possibility of line-of-sight (LoS) links, high mobility, on-demand deployment and low cost~\cite{9299862}. Several recent works such as in~\cite{LoRa-Based,s20082396}, and~\cite{8730598} have proposed the use of UAVs in the LoRa networks. In~\cite{LoRa-Based}, the authors implemented a prototype of LoRa-based air quality sensor on a UAV and a web-UI for user to configure the route of UAV and view the sensed data immediately. In~\cite{s20082396}, a UAV-assisted LoRa architecture is suggested in which UAVs act as relays for the traffic generated between LoRa nodes and a base station (BS). Then, they focus on designing a distributed topology control algorithm that periodically updates the UAV topology to adapt to the movement of the ground-based LoRa nodes. The work in~\cite{8730598} proposed measuring the RSSI at a LoRa gateway for indoor, suburban, and urban areas, when the LoRa transmitter is in another indoor location or mounted on a UAV. Their result shows that for a suburban environment, the drone height and antenna orientation have a crucial impact on the RSSI. Specifically, if the transmitting antenna is vertical, a stronger signal is received. In\cite{9827258}, the authors experimentally analyzed and modeled the channel of the UAV-to-ground LoRa links in the urban environments. Then, they discussed the dependencies between transmission power, spread factor (SF), RSSI, and signal-to-noise (SNR). However, the prior art in~\cite{LoRa-Based,s20082396} and~\cite{8730598} did not apply a practical localization method for a UAV-Assisted LoRa Networks, namely when used  for SAR operations in the highly remote areas.

The works in~\cite{two,9769039}, and~\cite{9795986} investigated the use of wearable LoRa radios to foster SAR missions in mountain environments. In~\cite{two}, the authors designed a localization system for SAR operations using LoRa. In this regard, they have characterized the path loss of a LoRa channel in mountain scenarios. However, the work in~\cite{two} mainly focused on LoRa channel modeling in a specific scenario without considering a flying LoRa gateway and UAVs. Moreover, the authors in~\cite{two} did not propose a general adaptive solution for SAR in new unknown environment. In~\cite{9769039} and~\cite{9795986}, the authors reported the measurements of the excessive aerial path loss for modeling ground-to-UAV links in a real mountain canyon, involving a receiving UAV and a transmitting LoRa radio worn by a volunteer lying on the rocks. They also demonstrated that LoRa radio propagation in the canyon is season-independent. Consequently, it is highly essential to practically design and analyze a localization system that leveraged a UAV-assisted LoRa network for the SAR operation in particular for highly remote areas. This is due to the fact that the ground-to-UAV LoRa link has less path loss compare to the ground-to-ground LoRa link. Moreover, by using the UAV as an FL gateway, it is possible to move the location of gateway in the sky in a quicker and more flexible way, compared to ground scenarios.

The main contribution of this paper is the implementation and analysis of a novel artificial intelligence-empowered search-and-rescue operation framework using a UAV-assisted LoRa network that can be applied to different unknown search environments. The proposed approach autonomously adapts the control policy of the UAV trajectory to the spatial geometry of a new search environment thereby allowing the system to determine the unknown location of a lost person. To solve the problem of the FL gateway control in the search-and-rescue system using the UAV-assisted LoRa network, we formulate a stochastic optimization problem whose goal is to maximize an episodic return that includes the received power from LoRa node at lost person over future time slots. Next, we model the FL gateway control problem as a partially observable Markov decision process (POMDP). Then, a deep reinforcement learning (RL) policy is proposed to adaptively control the FL gateway trajectory during SAR operation in a given environment. To find a near optimal solution, a parametric functional form policy is implemented using a deep recurrent neural network (RNN) that can directly search the optimal policies of the FL gateway controllers. Then, to increase the generalizability of the our framework, a control policy using deep meta-RL is designed. By applying deep meta-RL, the controller can integrate the prior FL gateway experience with information collected from the other search environments to train a rapidly adaptive policy model for SAR operation in a new environment~\cite{Mingje}. To analyze the performance of our proposed framework in the real world, we have experimentally designed and implemented our UAV-assisted LoRa networks including the LoRa end node as well as the FL and ground LoRa (GL) gateways. Then, we have done extensive experiments at Mongasht mountain ranges, near the areas of Ghaletol city, Khuzestan province, Iran. We have practically tested our SAR system in two different target areas: a wide plain and a slotted canyon. Practical evaluation results show that the FL gateway hovers over lost person's location after 50 and 141 time slots under deep meta-RL and RL control policy, respectively. Moreover, the average distance between UAV trajectories under deep meta-RL and deep RL based policies from the UAV trajectory under optimal policy are $619$ and $1930$ meter during the SAR operation time.

The rest of the paper is organized as follows. Section II describes system model and problem formulation. Section III proposes deep meta-RL framework for UAV gateway controlling in different unknown SAR environment. In Section~\ref{Sec:Exp_setup}, we introduce our experimental setup including the hardware that we have used to implement our UAV-Assisted LoRa Network and also our measurement scenarios. Then, in Section~\ref{Sec:Sim_setup}, we numerically evaluate the performance of our SAR system for highly remote areas and our proposed deep meta-RL-based UAV control policy which is trained by real data. Finally, conclusions are drawn in Section~\ref{Sec:Conclusion}.
\section{System model and problem formulation}
\subsection{System Model}
Consider a UAV-Assisted LoRa network composed of a LoRa node, an FL gateway, and one GL gateway.  Here, a lost person is equipped with a LoRa node that periodically transmits a known signal called beacon with duration $\tau$. The transmission power of the LoRa node is $P_{Tx}$ in dB. The location of the lost person is an unknown point of interest (POI) $(x_P,y_P)$ in the search area of interest (SAI), $\mathcal{C} \subset \mathbb{R}^2$.

The FL gateway is a LoRa gateway mounted on a UAV. The FL gateway is equipped with GPS and LoRa modules. At each time slot $t$, the FL gateway transmits a message, $\boldsymbol{m}_{t}=[\beta_{t},\gamma_{t},x_{t},y_{t}]$  to the GL gateway. This message contains the RSSI $\beta_t$ and SNR $\gamma_t$ of received LoRa beacon signal from LoRa node, as well as the FL gateway location, $(x_{t},y_{t})$. For simplicity, we assume the UAV is at the same height $z$ with the same speed $v$ all the time. In our model, the control action of  UAV is $a_{t}\in\{E,W,N,S,H\}$ where $E$, $W$, $N$, $S$, and $H$ represent the movement actions across the four cardinal points, i.e., east, west, north, south, and also hover on the current location. For example, if $a_{t}=N$, then in the next time slot $t+1$ and during $\tau$, the FL gateway will be at $(x_{t},y_{(t+1)})=(x_{t},y_{t}+v\tau)$. Following the received data of $\beta_{t},\gamma_{t}$ in the message $\boldsymbol{m}_{t}$ at the LoRa station, the received signal power $P_{Rx,t}$ at FL gateway and time slot $t$ can be computed as~\cite{Schwartz}:
\begin{align}
& P_{Rx,t}=\beta_t-10\log_{10}(1+10^{-\frac{\gamma_{t}}{10}}).\label{PRx}
\end{align}

The resulting received power $P_{Rx,t}$ at the FL gateway from unknown location of LoRa node is a random variable. This is due to the fact that the LoRa signals transmitting from the LoRa node antenna will often encounter the spatial geometry of SAI including random obstacles, such as trees and rocks, before reaching a given moving FL gateway receiver. The radiating electromagnetic field is reflected, diffracted, and scattered by these various obstacles, resulting commonly in a random multiplicity of rays impinging on the FL gateway antenna~\cite{Schwartz}.Thus, the radio geometry of a given SAI $i$ is directly affected by the spatial geometry. Generally, the statistically varying received signal power $P_{Rx,t}$ over wireless link between mobile FL gateway and LoRa node is modeled as follow~\cite{Schwartz}:
\begin{align}
& P_{Rx,t}=10\log_{10}{\nu^2}+\omega+10\log_{10}g(d_t)+P_{Tx}+  \nonumber \\
& 10\log_{10}(G_{Tx}G_{Rx}).\label{PRx_model}
\end{align}
where $10\log_{10}g(d_t)+P_{Tx}+10\log_{10}(G_{Tx}G_{Rx})$ is the far-field average power $\bar{P}_{Rx,t}$. $\omega$ is the shadow-fading random variable due to the large obstacles. $\nu$ is referred to multipath fading which is resulting from the rate of change of the signal being proportional to FL gateway velocity. Here, $d_{t}=\sqrt{(x_{t}-x_P)^2+(y_{t}-y_P)^2+z^2}$ represents the distance between the LoRa node carried by the flying gateway and the unknown location of lost person at time slot $t$. Given the random spatial geometry of each SAI, the resulting radio geometry parameters such as $\omega$, $\nu$, and function $g$ are unknown. In our model, we consider the worst case scenario in which there is no available model for the radio geometry parameters because the SAI is generally unknown for SAR operations. However, $g$ is a decreasing function with respect to $d_{t}$ in a UAV-assisted LoRa networks~\cite{Schwartz}. Then, we define a view circle centered at FL gateway with radius $d_{t}$ as follow:
\begin{align}
& \mathcal{C}_{t}=\{(x,y)| {(x_{t}-x_P)^2+(y_{t}-y_P)^2+z^2}\leq d_{t}^2  \}.\label{possible locations}
\end{align}
Indeed, from the point of view of the FL gateway, the possible location of the lost person is on the edge of this view circle $\mathcal{C}_{t}$ at time slot $t$. Since function $g$ decreases with the distance $d$, if the FL gateway moves toward lost person correctly, the radius of this circle decreases.

\begin{figure}[!t]
	\vspace{-0.5cm}
	\centering
	\includegraphics[width=8.5cm]{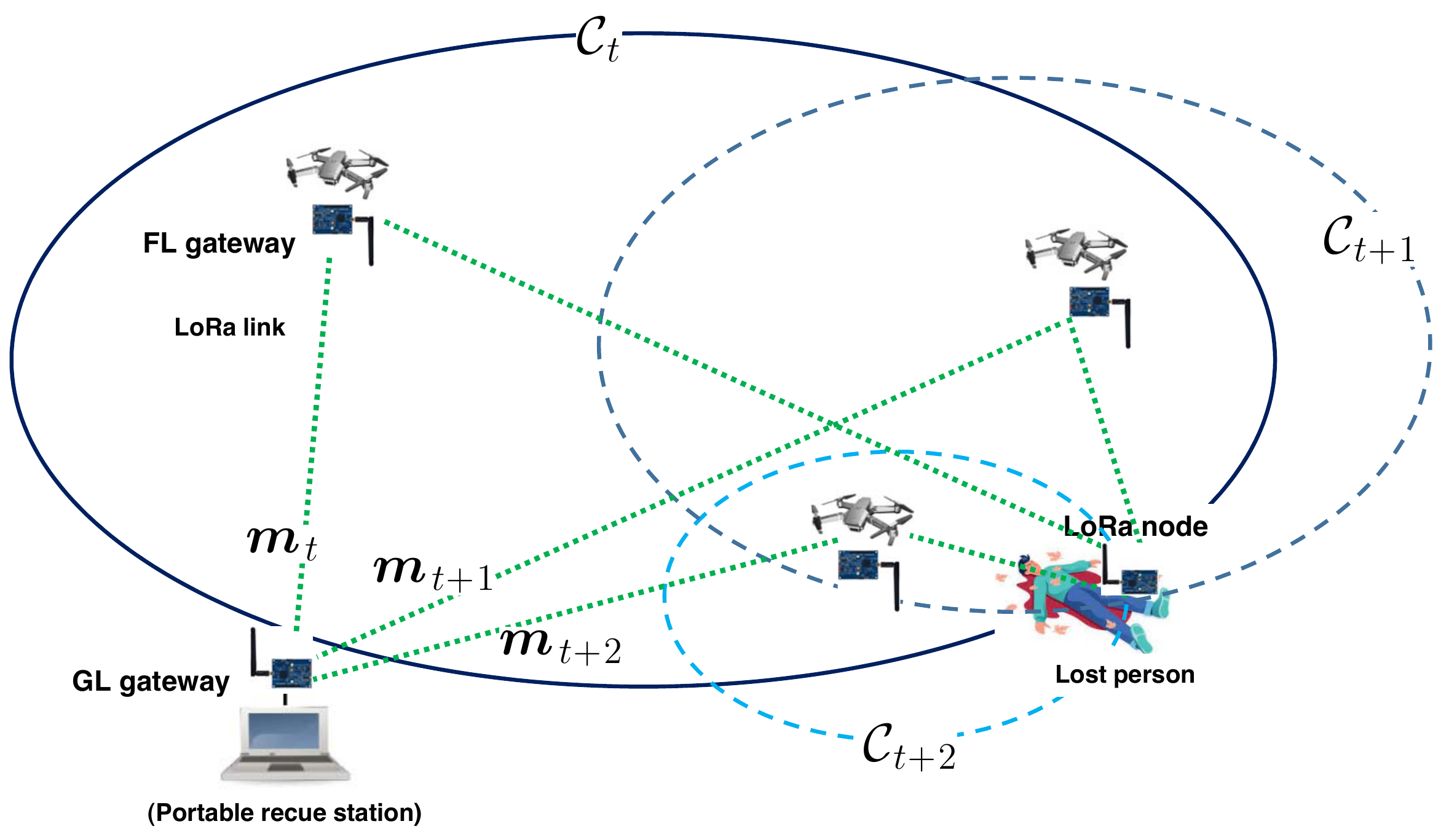}
	\caption{An illustrative example of the system model.}
	\label{System_Model1}
	\vspace{0.0cm}
\end{figure}

Fig.~\ref{System_Model1} illustrates our smart SAR system using UAV-assisted LoRa networks during three consecutive time slots $t$, $t + 1$, and $t + 2$. During these time slots, the portable rescue station equipped with GL gateway receives three messages $\boldsymbol{m}_{t}$, $\boldsymbol{m}_{t+1}$, and $\boldsymbol{m}_{t+2}$. As we can see in Fig.~\ref{System_Model1}, FL gateway has the view circles of $\mathcal{C}_{t}$, $\mathcal{C}_{t+1}$, and $\mathcal{C}_{t+2}$ at time slots $t$, $t+1$, and $t+2$. Following the received message from FL gateway, the possible locations of lost person will be in the edges of these view circles. As we can see in Fig.~\ref{System_Model1}, using the FL gateway control algorithm, the FL gateway moves toward the direction to increase the received power $P_{Rx,t}$ during time slots, $P_{Rx,t}<P_{Rx,t+1}<P_{Rx,t+2}$. Thus, the FL gateway moves toward the unknown location of the lost person during time slots. Note that, during the SAR operation, the GL gateway at the portable rescue station receives data messages from FL gateways over LoRa links. Thus, the FL gateway formation algorithm is run in the portable rescue station and all the SAR operation is monitored at the portable rescue in the real-time manner. Considering the stochastic changes in the received power at the FL gateway, designing an FL gateway control policy to move the UAV toward the lost person location is highly challenging particularly for different SAI scenarios with unknown radio geometry.

\subsection{Problem formulation}
Our goal is to characterize the FL gateway control policy which moves the UAV toward lost person over a future finite horizon $\mathcal{T}_t=\{t'|t'=t+1,...,t+T\}$ of length $T$ time slots. The objective of this policy is to minimize the set size of lost person possible location, $|\mathcal{C}_t|$, over ground of 2D dimension. Following (\ref{PRx_model}) and (\ref{possible locations}), and since $g$ is a decreasing function with respect to $d_{t}$ in UAV-assisted LoRa networks, minimizing the set size of lost person possible location, $|\mathcal{C}_t|$, is equivalent to increasing the received power, $P_{Rx,t}$, during the corresponding time slots. The FL gateway control policy at a given slot $t$ depends on the unknown radio geometry of SAI, which is a consequence of the stochastic nature of the wireless channel. Formally, we define a policy $\Pi_t=\{a_{t'}| \forall t' \in \mathcal{T}_t\}$ for the controller that assigns the next location of FL gateway. Consequently, we formulate the FL gateway control problem in our SAR system as follows:
\begin{align}
& \underset{\left\{
\Pi_t
\right\}}
\max \sum_{t'=t+1}^{t+T}\delta^{(t'-t)} P_{Rx,t'}, \label{Opt_prob}\\
& \hspace{0.1in} \text{s.t.}    \nonumber \\
& \hspace{0.1in} a_{t'}\in\{E,W,N,S,H\}, \forall t' \in \mathcal{T}_t, \label{Opt_prob_c1}
\end{align}
here,  $\delta$ is a discount factor. Maximizing the objective function in (\ref{Opt_prob}) ensures that the received power at FL gateway increases while the view circle surface of the FL gateway decreases. Thus, the FL gateways move toward the lost person during the considered time slots.

In practice, the solution of (\ref{Opt_prob}) faces the following challenges. First, since the location of the lost person is unknown, it is difficult to obtain the closed expression of the objective function in (\ref{Opt_prob}). Second, the received power distribution is a stochastic variable because the radio geometry wireless channels is dynamic and unknown. The complexity of the stochastic optimization problem in (\ref{Opt_prob}) becomes more significant due to the unknown probabilities for possible random network changes such as the fading over LoRa links and the user's location. Thus, the FL gateway control problem in (\ref{Opt_prob}) is a stochastic optimization problem that does not admit a closed-form solution and has an exponential complexity~\cite{Sutton2018}. Therefore, we propose a framework based on principles of the deep meta-RL for SAR operation in different unknown SAIs to solve the optimization problem in (\ref{Opt_prob}) with low complexity and in an adaptive manner. The proposed deep meta-RL method for FL gateway formation in UAV-assisted LoRa network, only takes UAV initial position, the RSSI and SNR of received LoRa beacon signal as input, then outputs the UAV trajectories after several episodes to move the UAV toward location of lost person.

\section{Deep Meta-Reinforcement Learning for SAR Operation}
In this section, we present the proposed adaptive control policy based on a deep meta-RL framework to solve the the FL gateway control problem in (\ref{Opt_prob}). Traditional policy gradient-based RL algorithms can only determine the adaptive FL gateway control policy in a fixed SAI including a fixed radio geometry. However, the meta-RL framework~\cite{refmeta-RL} is a novel learning approach that can integrate the prior FL gateway experience with information collected from the other SAI radio geometry to train a rapidly adaptive policy model for SAR operation a new SAI. Therefore, the proposed deep meta-RL can obtain the FL gateway control policies that can be quickly updated to adapt to new radio geometry properties using only a few further training steps. Next, we first introduce the deep RL algorithm for an adaptive FL gateway control policy in a given environment. Then, we explain the framework of deep meta-RL algorithm to train a rapidly adaptive policy model for a new SAI using the previous information collected from the given environment.
\subsection{Deep RL frame work for a given environment}
We model the problem in (\ref{Opt_prob}) as a partially observable Markov decision process (POMDP) represented by the tuple $\{\mathcal{S},\mathcal{A},\mathcal{O},P,R,o_0\}$. where $\mathcal{S}$ is the state space, $\mathcal{A}$ is the action space, $\mathcal{O}$ is the observation space, $P$ is the stochastic state transition function, $P(s',s,a) = \Pr(s_{t+1}= s'|s_t = s, a_t = a)$, $R_t(a_t,s_t)$ is the immediate reward function, and $o_0$ is the initial observation for the controller of the UAV to move FL gatewy\cite{threeRL}. Following POMDP, the required components of our proposed framework for a given SAI $\mathcal{C}$ are specified as follows:
\begin{itemize}
\item \emph{Agent:} the controller of UAV that moves the LoRa FL gateway.
\item \emph{Actions:} the control action of the agent at each time slot $t$ is a $a_{t}\in \mathcal{A}$. And, the action space $\mathcal{A}=\{E,W,N,S,H\}$ is the set of all optional actions including move east, west, north, south, and hover.
\item \emph{Observations:} the observation at time slot $t$ is the RSSI and SNR of received LoRa beacon signal by the FL gateway, and also current location of UAV, which are received with massage $\boldsymbol{m}_t$. Thus, $\boldsymbol{o}_{t}=[\beta_{t},\gamma_{t},x_{t},y_{t}]$, where $(x_{t},y_{t}) \in \mathcal{C}$. The observation space $\mathcal{O}$ is the set of all possible observations.
\item \emph{States:} the state at time slot $t$ is the  radio geometry characteristics including shadow-fading random variable $\omega$, multipath fading random variable $\nu$, unknown decreasing function $g$ of LoRa links between FL gateways and the LoRa node which is not observable due to the unknown location of lost person. In the case of POMDP, we consider the observation history of during $H$-consecutive previous time slots as state~\cite{three,Sutton2018}. Hence, the state at time slot $t$ in the environment $i$ is $\mathcal{H}_{t}=\cup_{h=0}^{H-1} \{\beta_{t-h},\gamma_{t-h},a_{t-h}\}$ containing the RSSI and SNR of received LoRa beacon signal and  the FL gateway control action at time slot $t-h$. The state space $\mathcal{S}$ is the set of all possible histories.
\item \emph{Immediate reward:} if the distance between the LoRa gateway and the LoRa node decreases, the received power at the LoRa gateway increases. Thus, we define the immediate reward as the power that FL gateway at time slot $t$, $R_{t}=P_{\text{Rx},t}$ which is given by (\ref{PRx}).
\item \emph{Episodic return:} if $\Lambda_{t}= \cup_{\forall t' \in \mathcal{T}_t} \{a_{t'},\boldsymbol{o}_{t'}\}$ is a trajectory of the POMDP during future $T$-consecutive time slots, then the stochastic episodic reward function during future $T$-consecutive time slots is defined as $R_{T,t}=\sum_{t'=t+1}^{t+T}R_{t}$.
\item  \emph{Policy:} for a given state, the policy is defined as the probability of the agent choosing each action. Our framework uses a functional-form policy parameterized by vector $\boldsymbol{\theta}$ to map the input state to the output action. Hence, the policy is expressed as $\pi_{\boldsymbol{\theta}}(a_{t},\mathcal{H}_{t})=\Pr(a_{t}|\mathcal{H}_{t})$
\end{itemize}

\begin{figure}[!t]
	\begin{center}
		\includegraphics[width=9cm]{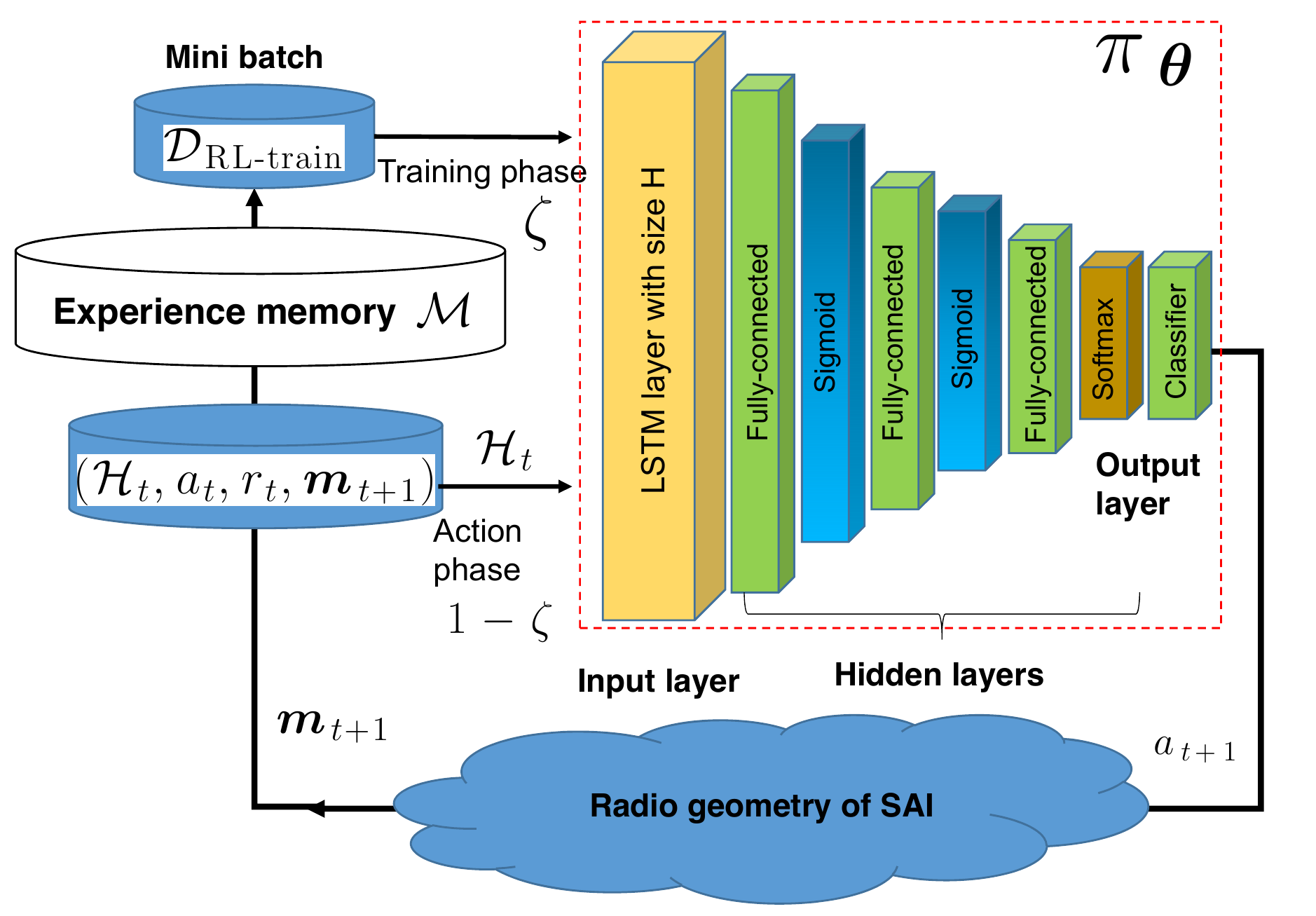}		\vspace{-0.4cm}
		\caption{ \small The Deep RL framework for implementing the FL gatway control policy $\pi_{\boldsymbol{\theta}}$.}                                    \vspace{-1.05cm}
		\label{System_model_RL}
	\end{center}
\end{figure}

The purpose of deep RL is to find the optimal policy that maximizes episodic return at FL gateway of UAV-assisted LoRa networks. Given the policy $\pi_{\boldsymbol{\theta}}$ and stochastic changes over the wireless LoRa link, the unknown probability of trajectory $\Lambda_{t}$ is equal to $\Pr(\Lambda_{t},\boldsymbol{\theta})=\prod_{\forall t' \in \mathcal{T}_t}\pi_{\boldsymbol{\theta}}(a_{t'},\mathcal{H}_{t'})\Pr\{\boldsymbol{o}_{(t'+1)}|a_{t'},\boldsymbol{o}_{t'}\}$ during future $T$-consecutive time periods. For a given SAI, we define the average episodic return for parameter vector $\boldsymbol{\theta}$ at time slot $t$ as $J_{t}(\boldsymbol{\theta}) =\sum_{\forall \Lambda_{t}} \Pr(\Lambda_{t},\boldsymbol{\theta}) R_{T,t}$. Given the parametric functional-form policy $\pi_{\boldsymbol{\theta}}$, the goal of the FL gateway controller is to solve the following optimization problem:
 \begin{align}
& \underset{\left\{
\substack{\boldsymbol{\theta} \in \mathbb{R}^{N}}
\right\}}
\max J_{t}(\boldsymbol{\theta}), \label{Opt_prob_RL} \\
& \hspace{0.1in} \text{s.t.}    \nonumber \\
& \hspace{0.1in} 0\leq \pi_{\boldsymbol{\theta}}(a_{t'},\mathcal{H}_{t'}) \leq 1, \forall a_{t'} \in \mathcal{A},\forall t' \in \mathcal{T}_t,\\
& \hspace{0.1in} \sum_{\forall a_{t'} \in \mathcal{A}} \pi_{\boldsymbol{\theta}}(a_{t'},\mathcal{H}_{t'}) = 1, \forall t' \in \mathcal{T}_t,
\end{align}
where $T<<N$ and $N$ is the number of parameters in the parametric functional-form policy $\pi_{\boldsymbol{\theta}}$. To solve the optimization problem in (\ref{Opt_prob_RL}), the FL gateway controller must have full knowledge about the transition probability $\Pr(\Lambda_{t},\boldsymbol{\theta})$, and all possible values of $R_{T,t}$ for all of the trajectories $\Lambda_{t}$ of the POMDP under policy $\pi_{\boldsymbol{\theta}}$. However, achieving this knowledge is not feasible, especially for a dynamic LoRa wireless channels between mobile FL gate way and LoRa node located at unknown location. To overcome this challenge, we propose to combine deep neural network (DNN) with the policy gradient-based RL method. Such a combination was shown in~\cite{three}, where a DNN learns a mapping from the partially observed state to an action without requiring any lookup table of all trajectories of observation values and policies over time. Consequently, we use a deep RL algorithm that includes a DNN to approximate the policy $\pi_{\boldsymbol{\theta}}$ for solving (\ref{Opt_prob_RL}). Our proposed DNN for deep RL method is presented in Fig.~\ref{System_model_RL}.  Here, the parameters $\boldsymbol{\theta} \in \mathbb{R}^{N}$ includes the weights over all connections of the proposed DNN where $N$ is equal to the number of connections~\cite{three}. The layers of the proposed deep NN for the implementing policy $\pi_{\boldsymbol{\theta}}$ are defined as follows:
\begin{itemize}
\item \emph{Input layer:} the input of proposed deep RL policy at time slot $t$ is the history of the POMDP during $H$-consecutive previous time slots, $\mathcal{H}_{t}$. Unlike the traditional DNN layers, we use a long short-term memory (LSTM) layer with size $H$ in the input of our policy DNN. This LSTM layer is an RNN that learns long-term dependencies between time steps in sequence data history of the POMDP trajectories. This is due to the fact that, in our model, the wireless channel affecting POMDP state transitions continuously depends on the spatiotemporal locations of UAV and the radio geometry of the SAR operation geographical area. Thus, we need to use deep RL that aggregates the observations over wireless links during previous time slots of the UAV trajectory and makes a more precise prediction of the next state of the POMDP~\cite{Goodfellow-2016}. Indeed, we use  LSTM layer in policy function to persist the hidden RSSI and SNR states across previous FL gateway trajectory for continued adaption to the radio geometry of a given environment.

\item \emph{Hidden layer:} the hidden layers include fully connected and Sigmoid layers. A sigmoid layer applies a Sigmoid function to the input such that the output is bounded in the interval (0,1). A fully connected layer multiplies the input by a weight matrix and then adds a bias vector.

\item \emph{Output layer:} the output layers include Softtmax and Classifier layers. The Softmax layer applies a Softmax function to the input. A classification layer computes the cross-entropy loss for classification and weighted classification tasks with mutually exclusive classes. In our model, the output shows the index of the actions in the the action space $\mathcal{A}$. More precisely, the output $\boldsymbol{y}_{t}=[y_{i,t}]\in \mathbb{R}^{T}$ is a vector of size $T$ actions where each element $i$ in this vector shows the action index of FL gateway at future time slot $t+i$.
\end{itemize}

The gradient of objective function in (\ref{Opt_prob_RL}) is $\nabla_{\boldsymbol{\theta}}J_{t}(\boldsymbol{\theta})= \sum_{\forall \Lambda_{t}} \nabla_{\boldsymbol{\theta}} \Pr(\Lambda_{t},\boldsymbol{\theta}) R_{T,t}$. Since $\nabla_{\boldsymbol{\theta}} \log \Pr(\Lambda_{t},\boldsymbol{\theta})=\frac{\nabla_{\boldsymbol{\theta}}\Pr(\Lambda_{t},\boldsymbol{\theta})}{\Pr(\Lambda_{t},\boldsymbol{\theta})}$, we can write  $\nabla_{\boldsymbol{\theta}}J_{t}(\boldsymbol{\theta})= \mathbb{E}_{\Lambda_{t}} \nabla_{\boldsymbol{\theta}} \log\Pr(\Lambda_{t},\boldsymbol{\theta}) R_{T,t}$.  Here,  $\Pr(\Lambda_{t},\boldsymbol{\theta})=\prod_{t'=t+1}^{t+T}\pi_{\boldsymbol{\theta}}(a_{t'},\mathcal{H}_{t'})\Pr\{\boldsymbol{o}_{((t'+1)}|\boldsymbol{o}_{t'},a_{t'}\}$ and $\nabla_{\boldsymbol{\theta}} \Pr\{\boldsymbol{o}_{((t'+1)}|\boldsymbol{o}_{t'},a_{t'}\}=0$. Thus, $\nabla_{\boldsymbol{\theta}}J_{t}(\boldsymbol{\theta})= \mathds{E}_{\Lambda_{t}}  \sum_{t'=t+1}^{t+T} \nabla_{\boldsymbol{\theta}} \log\pi_{\boldsymbol{\theta}}(a_{t'},\mathcal{H}_{t'})R_{T,t}$.  Having enough $M$ samples from trajectories $\Lambda_{t_m}$, one can approximate the expectation with sample-based estimator for $\nabla_{\boldsymbol{\theta}}J(\boldsymbol{\theta})$. As a result, we use the gradient-ascend algorithm to train the deep RL  policy $\pi_{\boldsymbol{\theta}}$  as follows:
\begin{equation*}
\nabla_{\boldsymbol{\theta}}J_t(\boldsymbol{\theta})\approx \frac{1}{M}\sum_{m=1}^M \Big(
\sum_{t'=t_m+1}^{t_m+T} \nabla_{\boldsymbol{\theta}} \log\pi_{\boldsymbol{\theta}}(a_{t'},\mathcal{H}_{t'}) R_{T,t'} \Big),
\end{equation*}
\begin{equation}
\boldsymbol{\theta}\leftarrow \boldsymbol{\theta}+\alpha_{\text{RL}} \nabla_{\boldsymbol{\theta}}J(\boldsymbol{\theta}), \label{RL-algorithm}
\end{equation}
where $\alpha_{\text{RL}}$ is the reinforcement learning rate. As shown in Fig.~\ref{System_model_RL}, the training batch set $\mathcal{D}_{\text{RL, train}}$ is randomly selected from experience memory $\mathcal{M}$. Thus, a batch of deep RL training set $\mathcal{D}_{\text{RL,\text{train}}}$ of $M$ samples is available enough to train the deep NN network of $\pi_{\boldsymbol{\theta}}$ during the time. Each training sample $m$ in $\mathcal{D}_{\text{RL,train}}$ includes histories during $H$-consecutive time slots before time slot $t_m$, $\mathcal{H}_{t_m}$, and actions in the trajectory of future $T$-consecutive time slots after time slot $t_m$, $\Lambda_{t_m}$. In summary, we implement the parametric functional-form policy $\pi_{\boldsymbol{\theta}}$ with the proposed deep NN in Fig.~\ref{System_model_RL}.  The proposed deep RL algorithm for FL gateway control is summarized in Algorithm~\ref{algorithm-RL}. During the SAR time, the deep RL policy is trained with probability $\xi$ based on the gradient-ascend algorithm in (\ref{RL-algorithm}). In the training phase, the deep RL policy is  adaptively trained using train data set from experience memory $\mathcal{M}$. In the update phase, the experience memory is updated with history and trajectories during the time slots. The FL gateway moves using this deep RL-based algorithm until the reward $R_{t}$ becomes more than defined target reward $R_{\text{target}}$. This means that the UAV is close enough to the lost person and the received power is more than $R_{\text{target}}$.

\begin{algorithm}[!t]
\caption{\small{Proposed deep RL-based algorithm for FL gateway control}} \label{algorithm-RL}
\small{
\begin{algorithmic}[1]
\State \textbf{Input:} Initial location UAV; RL learning rate $\alpha_{\text{RL}}$; training probability $\xi$; a defined deep NN for policy $\pi_{\boldsymbol{\theta}}$; initial value for $\boldsymbol{\theta}$; batch size $M$; target reward $R_{\text{target}}$.
\State \textbf{Online deep RL}
\Repeat
\State{\textbf{Exploiting phase}: apply deep RL control policy $\pi_{\boldsymbol{\theta}}$ depicted in Fig.~\ref{System_model_RL};}
\State {\textbf{Update phase}: update the experience memory set $\mathcal{M}=\mathcal{M} \cup \{\mathcal{H}_{t}\cup\Lambda_{t}\}$;}
\State{\textbf{Training phase}: with probability $\zeta$, train deep RL control policy $\pi_{\boldsymbol{\theta}}$ as follows:}
\parState {Uniformly select a set of $M$ minibatch samples from updated  experience memory set $\mathcal{M}$ as a training set $\mathcal{D}_{\text{RL, train}}$};
\parState {Train the deep RL control policy, $\pi_{\boldsymbol{\theta}}$, following the gradient-ascend algorithm in (\ref{RL-algorithm});}
\Until {$R_{t}>R_{\text{target}}$}
\State \textbf{Ouput:} adaptive control policy $\pi_{\boldsymbol{\theta}}$ during SAR operation time.
\end{algorithmic}
}
\end{algorithm}

The spatial geometry of a given environment affects on the channel fading due to the the the blockage, reflection, and refraction wireless waves. Thus, the radio geometry of a given environment is directly affected by the spatial geometry. However, unlike ground-to-ground wireless links, a UAV-to-ground LoRa link is more robust to the fading effects resulting from the spatial geometry. Due to this fact, compare to the ground-to-ground LoRa links, the radio geometry information resulting from UAV-to-ground LoRa link is more stable. Moreover, given the radio geometry information resulting from UAV-to-ground LoRa link, the knowledge gained while learning the FL gateway control policy in a given spacial geometry could be applied when trying to recognize a new FL gateway control policy in another spacial geometry. Consequently, next, we use the information of a trained deep RL policy of FL gateway control in a given environment to design of a control policy for a SAR operation in a new SAI.
\subsection{Deep meta-RL framework for new environments}
 We introduce the deep meta-RL framework to use the information from a given environment to design a control policy for a new unknown radio geometry. Compare to deep RL policy, the proposed deep meta-RL policy can integrate the prior experience in one environment with information collected from the FL gateway movement in a new search environment; thus, training a rapidly adaptive learning model for FL gateway control. Indeed, the meta training procedure requires a realization set of state and near optimal policy. Here, we use the realization set of states and actions in the success full SAR operation in different SAIs to design a policy for SAI environment. In this case, we define our task as follow:
\begin{itemize}
\item \emph{Tasks:} Given the history of POMDP, task $\mathcal{T}$ is the realization of FL control policy solve the optimization problem in (\ref{Opt_prob_RL}) in each time slot $t$ at each environment with specific radio geometry. Thus, for a given SAI environment, the $\mathcal{T}_{t_k}=\{\mathcal{H}_{t_k}\cup\Lambda_{t_k}\}$ include history and trajectory under control policy at time slot $t_k$ of SAR operation which is accessible from experience memory in Fig~\ref{System_model_RL}.
\item \emph{Meta-train dataset:} $\mathcal{D}_{\text{Meta-train}}=\cup_{k=1}^K \mathcal{T}_{t_k}$ is defined as $K$ different tasks in previous successful SAR operation.
\item \emph{Meta-test dataset:} $\mathcal{D}_{\text{Meta-test}}=\cup_{e=1}^E \{ \mathcal{H}_{t_e}\cup a_{t_e} \}$ is defined as $E$ different histories and actions in the new environment under target policy $\pi_{\boldsymbol{\psi},\boldsymbol{\phi}}$.
\end{itemize}

Here, we use the idea of the most popular policy-gradient meta-RL method which is called using model agnostic meta-learning (MAML) to design the FL gateway control policy in the new search area of interest using the previous successful SAR information~\cite{27}. During the meta training procedure, a Meta-train dataset $\mathcal{D}_{\text{Meta-train}}$ is first sampled from experienced memory of previous successful SAR operation. Then, the meta-RL method collects experience information in variable $\boldsymbol{z}$ from $\mathcal{D}_{\text{Meta-train}}$ and use the deep meta-RL policy function $\pi_{\boldsymbol{\phi},\boldsymbol{\psi}}$ to predict actions given the history $\mathcal{H}_{t}$ of the new environment. Indeed, the deep meta-RL policy function $\pi_{\boldsymbol{\phi},\boldsymbol{\psi}}=  \pi_{\boldsymbol{\psi}} \pi_{\boldsymbol{\phi},\boldsymbol{z}}$ in which $\pi_{\boldsymbol{\psi}}=\Pr(\boldsymbol{z})$ is the probability of experience information variable $\boldsymbol{z}$ and $\pi_{\boldsymbol{\phi},\boldsymbol{z}}=\Pr(a_{t}|\mathcal{H}_{t},\boldsymbol{z})$ is the probability of chosen action $a_{t}$ given the history $\mathcal{H}_{t}$ of the new environment and encoded experience information data $\boldsymbol{z}$ from expreince of the previous successful SAR operations. More concretely, the objective of the meta training procedure is as follows:
\begin{align}
& \underset{\left\{
\pi_{\boldsymbol{\phi},\boldsymbol{\psi}}
\right\}}
\max \text{ } J_{t}(\boldsymbol{\phi},\boldsymbol{\psi}), \label{Opt_MetaRL}\\
& \hspace{0.1in} \text{s.t.}    \nonumber \\
& \hspace{0.1in} 0\leq \pi_{\boldsymbol{\phi},\boldsymbol{\psi}}(a_{t'},\mathcal{H}_{t'}) \leq 1, \forall a_{t'} \in \mathcal{A},\forall t' \in \mathcal{T},\\
& \hspace{0.1in} \sum_{\forall a_{t'} \in \mathcal{A}} \pi_{\boldsymbol{\phi},\boldsymbol{\psi}}(a_{t'},\mathcal{H}_{t'}) = 1, \forall t' \in \mathcal{T},
\end{align}
where the parametric functional-form $\pi_{\boldsymbol{\psi}}$ encodes the experience information from tasks in Meta-train dataset $\mathcal{D}_{\text{Meta-train}}$ to help $\pi_{\boldsymbol{\phi},z}$ in finding optimal policy in the new environment. In our deep meta-RL framework, we use a DNN to approximate the policy $\pi_{\boldsymbol{\psi}}$, where the parameters $\psi$ includes the weights over all connections of the DNN.
\begin{figure}[!t]
	\begin{center}
		\includegraphics[width=9cm]{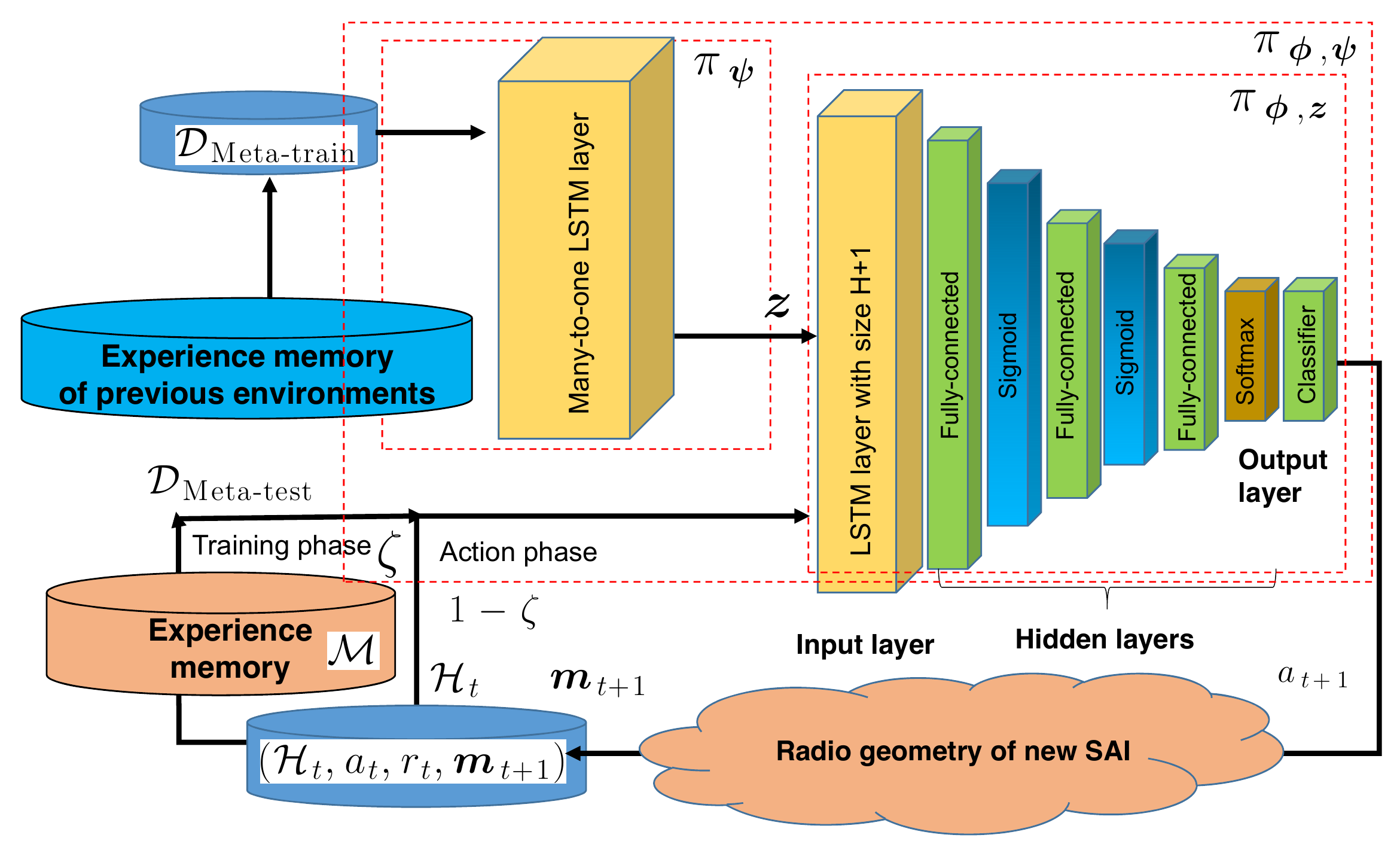}		\vspace{-0.4cm}
		\caption{ \small The deep meta-RL framework for implementing the FL gatway control policy $\pi_{\boldsymbol{\phi},\boldsymbol{\psi}}$.}                                    \vspace{-1.05cm}
		\label{System_model_Meta}
	\end{center}
\end{figure}

Having enough $M_1$ samples from trajectories $\Lambda_{t_{m_1}}$ in the dataset $\mathcal{D}_{\text{Meta-test}}$ from experienced memory in new SAI, one can approximate the expectation with sample-based estimator for $\nabla_{\boldsymbol{\phi,\psi}}J_{t}(\boldsymbol{\phi},\boldsymbol{\psi})$. Here, $\nabla_{\boldsymbol{\phi}} \log\pi_{\boldsymbol{\phi},\boldsymbol{\psi}}=\nabla_{\boldsymbol{\phi}} \log\pi_{\boldsymbol{\psi}} \pi_{\boldsymbol{\phi},\boldsymbol{z}}$ which is equal to $\nabla_{\boldsymbol{\phi}} \log\pi_{\boldsymbol{\psi}}+\nabla_{\boldsymbol{\phi}} \log \pi_{\boldsymbol{\phi},\boldsymbol{z}}$. Thus, for a given $\boldsymbol{\psi}_0$, $\boldsymbol{z}_0=\pi_{\boldsymbol{\psi}_0}(\mathcal{D}_{\text{Meta-train}})$ and  $\nabla_{\boldsymbol{\phi}} \log\pi_{\boldsymbol{\phi},\boldsymbol{\psi}_0}=\nabla_{\boldsymbol{\phi}} \log \pi_{\boldsymbol{\phi},\boldsymbol{z}_0}$. In this case, we use the gradient-ascend algorithm to train the deep meta-RL  policy $\pi_{\boldsymbol{\phi,\psi}}$ with respect to $\boldsymbol{\phi}$  as follows:
\begin{align}
&\nabla_{\boldsymbol{\phi}}J(\boldsymbol{\phi,\psi_0})\approx \nonumber \\
&\frac{1}{M_1}\sum_{m=1}^{M_1} \Big(
\sum_{t'=t_{m_1}+1}^{t_{m_1}+T} \nabla_{\boldsymbol{\phi}} \log\pi_{\boldsymbol{\phi},\boldsymbol{z}_0}(a_{t'},\mathcal{H}_{t'}) R_{T,t'} \Big),\nonumber \\
&\boldsymbol{z}_0=\pi_{\boldsymbol{\psi}_0}(\mathcal{D}_{\text{Meta-train}}),\nonumber \\
&\boldsymbol{\phi}\leftarrow \boldsymbol{\phi}+\alpha_{\text{Meta-RL},1} \nabla_{\boldsymbol{\phi}}J(\boldsymbol{\phi,\psi_0}). \label{Meta_RL-algorithm_1}
\end{align}
Since the parameters $\boldsymbol{\phi}$ are updated based on collected data in $\mathcal{D}_{\text{Meta-test}}$ of new environment, the phase of updating $\boldsymbol{\phi}$ is called adaption phase. Here, $\alpha_{\text{Meta-RL},1}$ is the meta-RL rate for the adaption phase.

For a given $\boldsymbol{\phi}_0$, we have $\nabla_{\boldsymbol{\psi}} \log\pi_{\boldsymbol{\phi}_0,\boldsymbol{\psi}}=\nabla_{\boldsymbol{\psi}} \log\pi_{\boldsymbol{\psi}}+\nabla_{\boldsymbol{\psi}} \log \pi_{\boldsymbol{\phi}_0,\boldsymbol{z}}$ which is equal to $\nabla_{\boldsymbol{\psi}} \log\pi_{\boldsymbol{\psi}}$.

Having enough $M_2$ samples from trajectories $\Lambda_{t_{m_2}}$ in the dataset $\mathcal{D}_{\text{Meta-train}}$ from experienced memory in previous successful SAR operations in different environments, we use the gradient-ascend algorithm to train the deep meta-RL  policy $\pi_{\boldsymbol{\phi_0,\psi}}$ with respect to $\boldsymbol{\psi}$  as follows:
\begin{align}
&\nabla_{\boldsymbol{\psi}}J(\boldsymbol{\phi}_0,\psi)\approx \nonumber \\
&\frac{1}{M_2}\sum_{m=1}^{M_2} \Big(
\sum_{t'=t_{m_2}+1}^{t_{m_2}+T} \nabla_{\boldsymbol{\psi}} \log\pi_{\boldsymbol{\psi}}(a_{t'},\mathcal{H}_{t'}) R_{T,t_{m_2}} \Big),\nonumber \\
&\boldsymbol{\psi}\leftarrow \boldsymbol{\psi}+\alpha_{\text{Meta-RL},2} \nabla_{\boldsymbol{\psi}}J(\boldsymbol{\phi}_0,\psi), \label{Meta_RL-algorithm_2}
\end{align}
where $\alpha_{\text{Meta-RL},2}$ is the meta-RL rate for Meta-train set.

As shown in Fig.~\ref{System_model_Meta}, the training set $\mathcal{D}_{\text{Meta, train}}$ is randomly selected from experience memory of previous successful SAR operations in different environments. However, each adaptation sample $m_1$ includes histories during $H$-consecutive time slots before time slot $t_{m_1}$, $\mathcal{H}_{t_{m_1}}$, and actions in the trajectory of future $T$-consecutive time slots after time slot $t_{m_1}$, $\Lambda_{t_{m_1}}$ in the new environment. In summary, we implement the parametric functional-form policy $\pi_{\boldsymbol{\phi},\boldsymbol{\psi}}$ with the proposed deep NN in Fig.~\ref{System_model_Meta}.  The proposed deep meta-RL algorithm for FL gateway control is summarized in Algorithm~\ref{algorithm-Meta-RL}. In the meta-training phase, the deep meta-Rl policy is trained using train data set from experience memory of successful SAR operations in previous environments, and in the adaptation phase, the experience memory including history and trajectories of the new environment during the time of SAR operation is used.
\begin{algorithm}[!t]
\caption{\small{Proposed deep meta-RL-based algorithm for FL gateway control in new environment}} \label{algorithm-Meta-RL}
\small{
\begin{algorithmic}[1]
\State \textbf{Input:} Initial location UAV; meta-RL learning rates $\alpha_{\text{Meta-RL},1}$ and $\alpha_{\text{Meta-RL},2}$; training probability $\zeta$; a defined deep NN for policy $\pi_{\boldsymbol{\phi},\boldsymbol{\psi}}$; initial value for $\boldsymbol{\phi}$ and $\boldsymbol{\psi}$; batch size $M$; experienced memory $\mathcal{M}_I$; target reward $R_{\text{target}}$
\State \textbf{Online deep meta-RL}
\Repeat
\State{\textbf{Exploiting phase}: apply deep meta-RL control policy $\pi_{\boldsymbol{\phi},\boldsymbol{\psi}}$  depicted in Fig.~\ref{System_model_Meta};}
\State {\textbf{Update phase}: update the experience memory $\mathcal{M}=\mathcal{M} \cup \{\mathcal{H}_{t}\cup\Lambda_{t}\}$;}
\State{\textbf{Training phase}: with probability $\xi$, train deep meta-RL control policy $\pi_{\boldsymbol{\phi},\boldsymbol{\psi}}$ as follows:}
\parState {Uniformly select a set of $M_2$ history and action samples from $\mathcal{M}_2$ as a adaption-training set $\mathcal{D}_{\text{Meta-train}}$};
\parState {Uniformly select a set of $K$ task samples $\mathcal{M}$ as a meta-training set $\mathcal{D}_{\text{Meta-train}}$};
\parState {Given meta-RL learning rates $\alpha_{\text{Meta-RL},1}$ and $\alpha_{\text{Meta-RL},2}$, train the deep meta-RL control policy, $\pi_{\boldsymbol{\phi},\boldsymbol{\psi}}$ , following the gradient-ascend algorithms in (\ref{Meta_RL-algorithm_1}) to update $\boldsymbol{\phi}$ and (\ref{Meta_RL-algorithm_2}) to update $\boldsymbol{\psi}$;}
\Until {$R_{T,t}>R_{\text{target}}$}
\State \textbf{Ouput:} adaptive control policy $\pi_{\boldsymbol{\phi},\boldsymbol{\psi}}$ during SAR time in the new environment.
\end{algorithmic}
}
\end{algorithm}

\section{Experimental Setup}\label{Sec:Exp_setup}
The our implementation of  the UAV-assisted LoRa networks and environments for our experimental testbed are presented and discussed in this section.
\subsection{Considered Hardware}
Our experimental setup of UAV-assisted LoRa networks includes the LoRa end node, FL and GL gateways. In our setup, all of the designed nodes, and gateways are powered using lithium polymer (Li-Po) batteries with output voltage of 3.7 V and capacity of 1100 mAh. We have designed our printed circuit boards (PCBs) to assemble LoRa node FL gateways on them. Using ATmega328 which is a single-chip microcontroller created by Atmel in the megaAVR family, the LoRa node FL gateways are assembled and programmed in C language ~\cite{1_Esetup}. In our setup, we use the LoRa Ra-02 module from Ai-Thinker which is equipped by a LoRa SX1278 Semtech core~\cite{2_Esetup} and~\cite{3_Esetup}.

\begin{figure}[t!]
\begin{subfigure}{.5\textwidth}
  \centering
	\includegraphics[height=4cm]{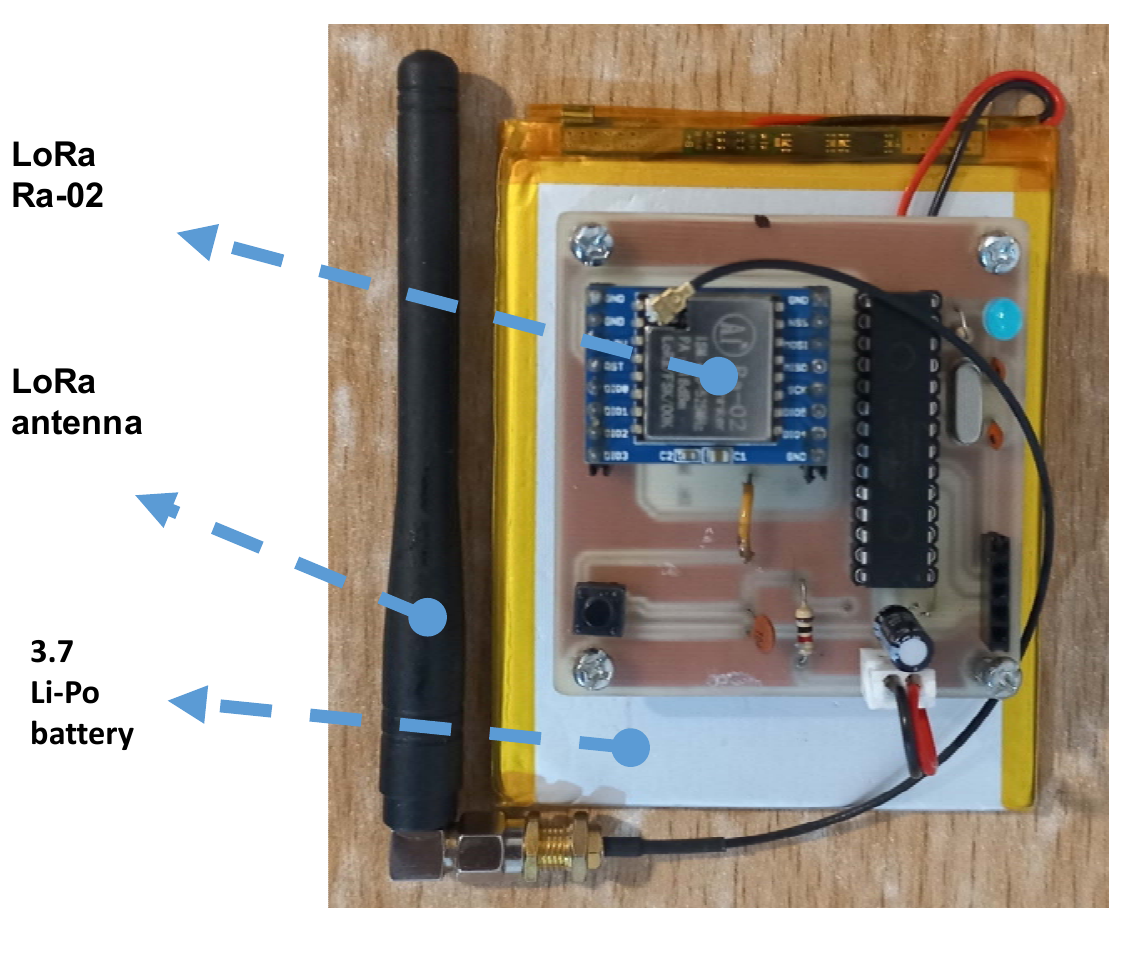}
	\caption{\small{Assembled LoRa end node.}}
  \label{Experimental Setup_node}
\end{subfigure}
\begin{subfigure}{.5\textwidth}
  \centering
 	\includegraphics[height=4cm]{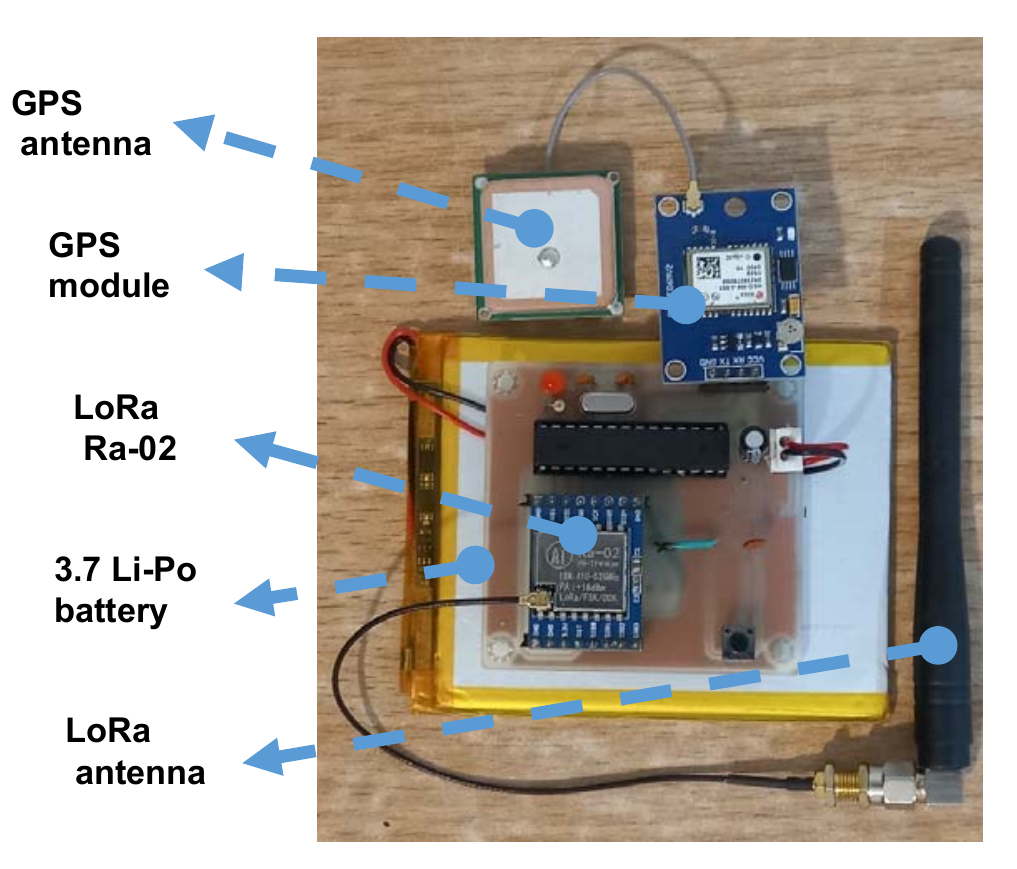}
	\caption{\small{Assembled Flying LoRa gateway node.}}
  \label{Experimental Setup_Fnode}
\end{subfigure}
\begin{subfigure}{.5\textwidth}
  \centering
	\includegraphics[height=4cm]{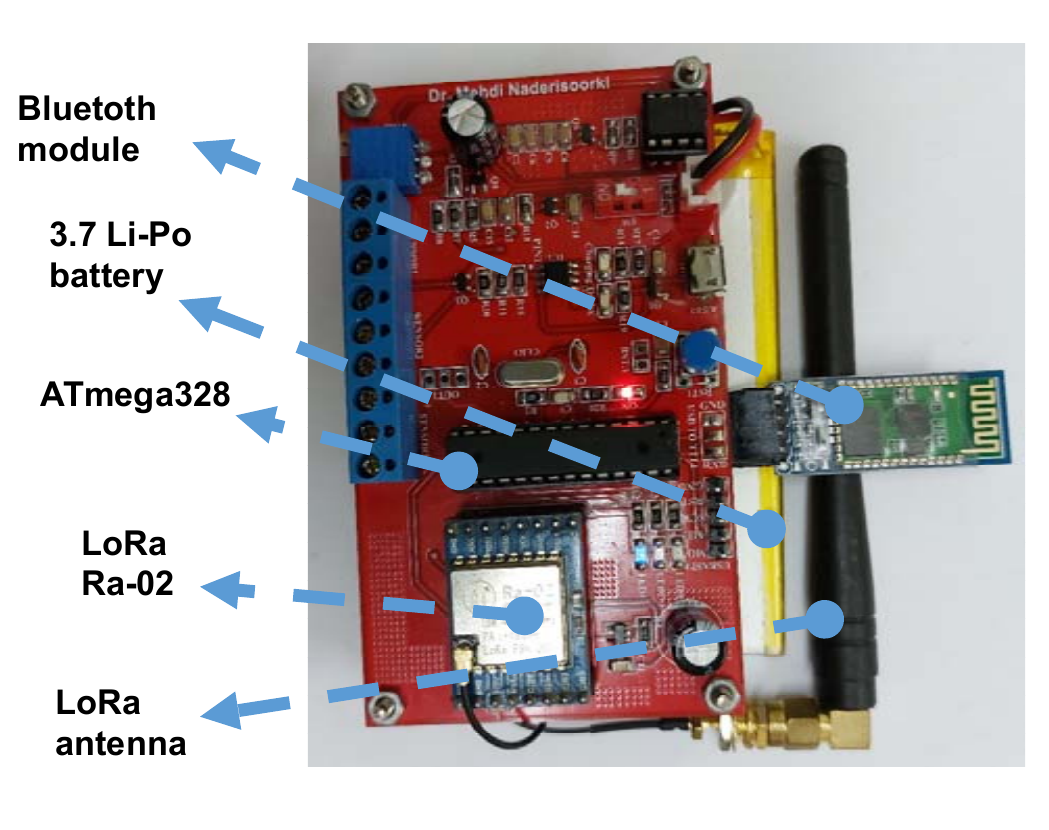}
	\caption{\small{Assembled ground LoRa gateway node.}}
  \label{Experimental Setup_Gnode}
\end{subfigure}
\caption{\small{Assembled wireless nodes of the UAV-Assisted LoRa Network.}}
\label{Experimental Setup}
\end{figure}

The designed LoRa node has a LoRa module which is connected to a commercial external folded dipole antenna, and transmits a beacon packet using LoRa module on the ISM frequency band at 433 MHz. The detail of LoRa node is shown in Fig.~\ref{Experimental Setup_node}. The FL gateway board is equipped with LoRa Ra-02 and GPS modules in which the LoRa Ra-02 module is connected to to 433 Mhz antenna and the GPS module has the square-shape receiver GPS antenna. The designed of FL gateway board is shown in Fig.~\ref{Experimental Setup_Fnode}. The dimensions of FL gateway board are 295 mm×60 mm×20 mm and its weight is 100 gr. Thus, the weight of designed FL gateway board is suitable to be mounted on the commercial drones and UAVs. We mount the FL gateway board over DJI Phantom 4 Pro to set up an FLoRa gateway. To use the maximum antenna radiation, we firmly fix the gateway node under DJI Phantom 4 Pro, while the LoRa antenna is vertically pointed toward the earth surface. In our setup, the FL gateway receives the beacon packet using LoRa module and GPS coordinations of UAV using GPS module. Then, The FL gateway retransmits a data packet containing of the RSSI and SNR of received LoRa beacon signal and GPS coordination of UAV over UAV-to-ground LoRa link to the be received by GL gateway.

The GL gateway consists of a LoRa Ra-02 module connected to 433Mhz antenna and our designed board. The head of rescue operation can connect to the Gl gateway vie USB cable or Bluetooth wireless link. The GL gateway setsup is shown in Fig.~\ref{Experimental Setup_Gnode}.The GL gateway receives data packets from FL gateways using LoRa module, and transmits these received packets to the portable computer such as Labtop over USB cable or to the mobile phone over Bluetooth link. In this case, the rescue operation can be monitored and analyzed the using the data that are gather on computer or mobile phone in a realtime manner. The realtime received data packets are used to run our proposed deep reinforcement algorithms. Following our proposed deep RL algorithms, the control policy suggest the next movement of UAV, and the pilot move the UAV toward that direction. Using the application which is written in Python, the trajectory of UAV is depicted for the rescue operation head. And the UAV moves Fl gateway toward the location of LoRa end node. Thus, the location of lost person is find very fast.

\subsection{Experimental Testbed}
\begin{figure}[ht]
\begin{subfigure}{.5\textwidth}
  \centering
  \includegraphics[width=8cm]{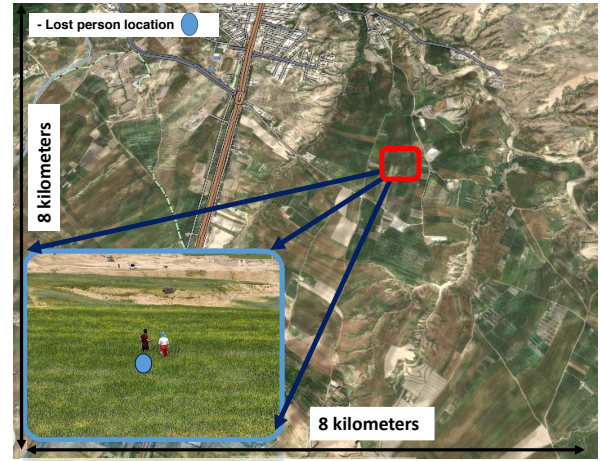}
  \caption{A very wide plain covered with 100cm-height  grasses near Qaletol city, Khuzestan province, Iran.}
  \label{Rate_Time_Dist}
\end{subfigure}
\begin{subfigure}{.5\textwidth}
  \centering
 \includegraphics[width=8cm]{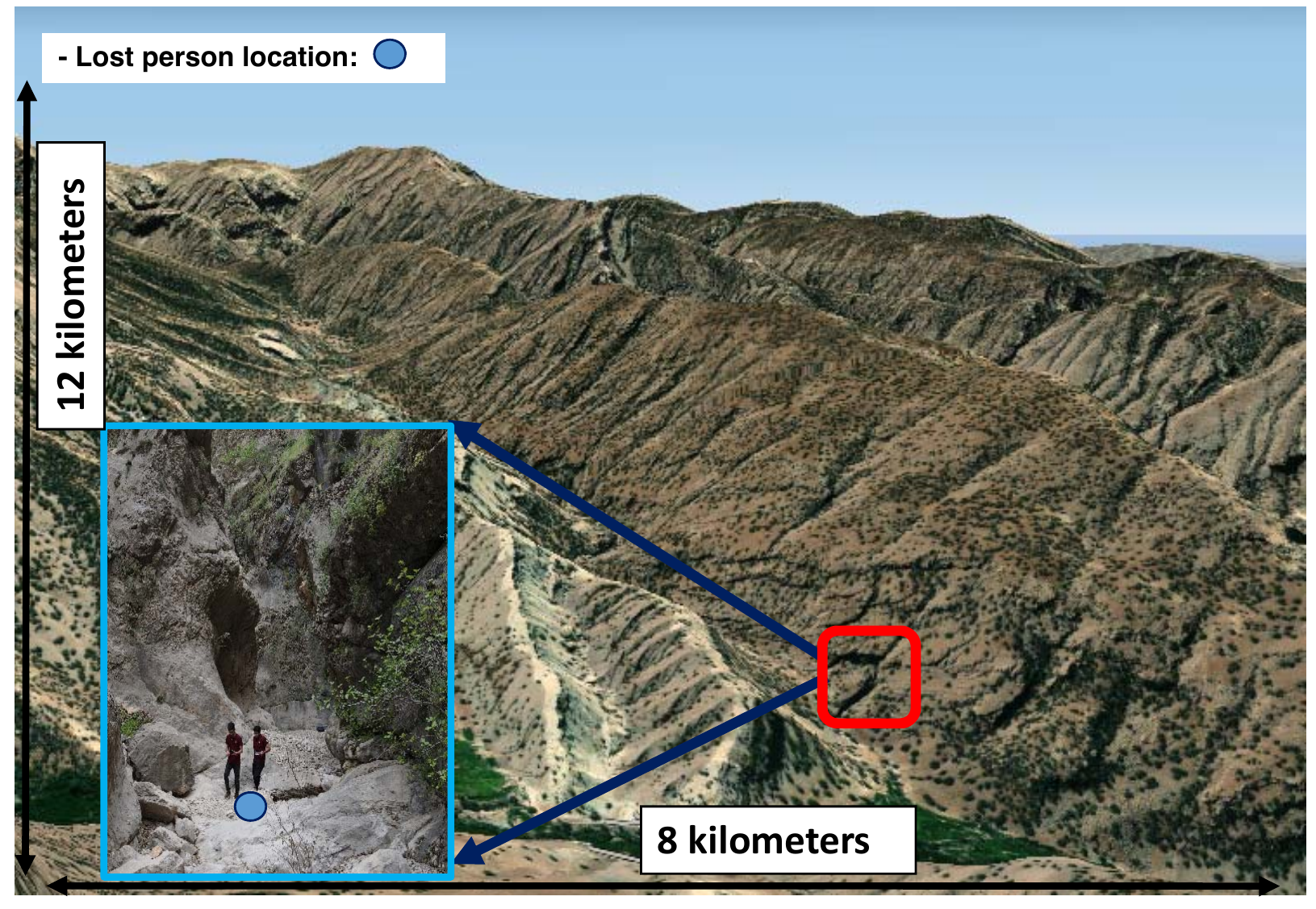}
  \caption{A slotted canyon with 150m-height walls in the Mongasht mountain ranges, Khuzestan province, Iran.}
  \label{Rate_Time_Cent}
\end{subfigure}
\caption{\small{Experimental Scenarios. These satellite images are taken from ``www.fatmap.com''.}}
\label{experimental Scenarios}
\end{figure}

The measurement is at Mongasht mountain ranges, near the areas of Ghaletol city, Khuzestan province, Iran. For the lost person, the LoRa node is held in the students’ hand while the students move to the unknown location in the target area. And for the FLoRa gateway, the gateway node is mounted on the UAV. The UAV hovers the gateway node at 300 meters. This elevation is high enough that UAV does not hit any obstacle such as mountains or trees on its path. Since, we are interested to evaluate our SAR system in the different radio geometry including both LoS and NLoS links, we have done our experiments on two different target areas: a wide plain and a slotted canyon. When the target area of lost person is a wide plain, there is a LoS path between GLoRa or FLoRa gateway and LoRa nodes. However, when the lost person is in a slotted canyon, there may not be any LoS path between GLoRa or FLoRa gateway and nodes, and the LoRa nodes are in the shadowing area of mountain walls around the slotted canyon. In Fig.~\ref{experimental Scenarios}, the experimental scenarios have been shown which are plain and canyon scenarios.

\section{Performance analysis}\label{Sec:Sim_setup}
In this section, we evaluate the performance of our proposed  deep meta-reinforcement learning approach for SAR operation using the real measurement of UAV-assited LoRa network for the scenarios at Mongasht mountain ranges. In our measurement, the duration of each time slots is 2 seconds, the UAV speed is $20$ meter per second. The maximum  life time of one UAV battery is $20$ minutes. We compare deep learning policy  with optimal and greedy policies as the benchmarks. In the optimal policy, we give the unknown location of lost person to the UAV pilot, then, the UAV directly moves toward target location. Under the greedy policy, there tow phases: sense and action. The UAV starts the sense phase with probability of $0.1$, in which UAV sequentially moves north, east, south, and west for 3 time slots to measure received power. The UAV starts action phase with probability of $0.9$, in which UAV moves toward the direction with highest received power at the previous sense phase.

\begin{figure}[t!]
  \centering
 \includegraphics[width=8cm]{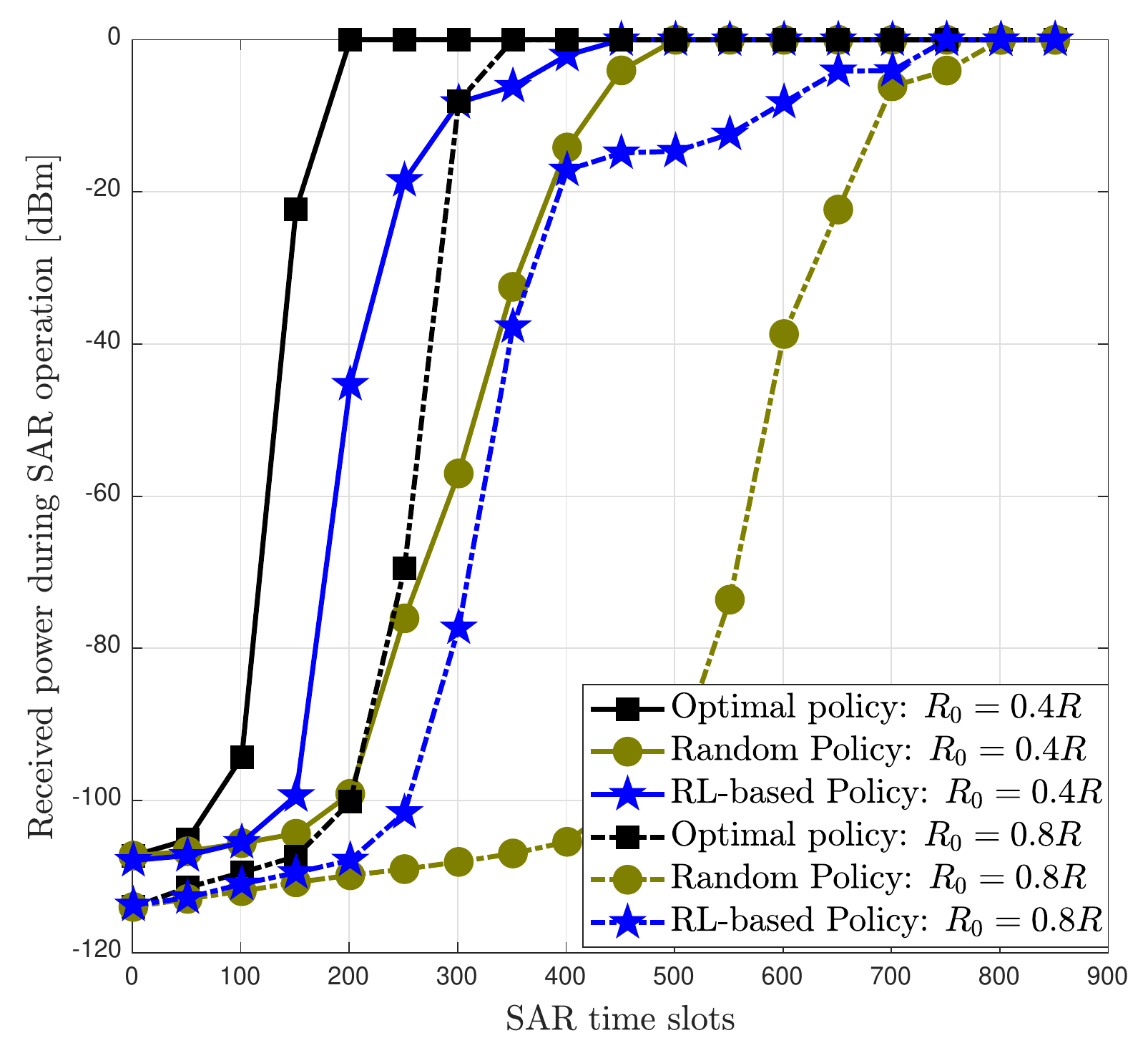}
 \caption{Received power at FL gateway v.s. rescue time slots in wide plain environment.}
   \label{RL_Plain_RSSI}
   \vspace{-5mm}
\end{figure}

In Fig.~\ref{RL_Plain_RSSI}, we show the received power at FL gateway versus rescue time slots in the wide plain environment. From Fig.~\ref{RL_Plain_RSSI}, we observe that the received power at FL gateway increases in less time slots under the optimal policy. However, when the greedy algorithm is used, the received power at FL gateway increases slower. The performance of proposed deep learning policy is between optimal and greedy policies. As we can see in Fig.~\ref{RL_Plain_RSSI}, when the UAV initial location is $0.8R$ far from lost person, the received power at FL gateway received to its maximum value around SAR time slot 300, 700, 800  under the optimal, deep RL, and greedy algorithms, respectively. Moreover, in the plain environment, the greedy and deep learning policies converge and the UAV can find the lost person finally because in the plain environment there are mostly LoS link between lost person and FL gateway. under deep learning policy, when we start the initial location of UAV SAR operation at $0.4$ radius of plain environment the UAV can find the lost person at time slot 400, however is the UAV initial location is at $0.8$ radius of plain environment, the lost person is found at time slot 800.

\begin{figure}[t!]
  \centering
 \includegraphics[width=8cm]{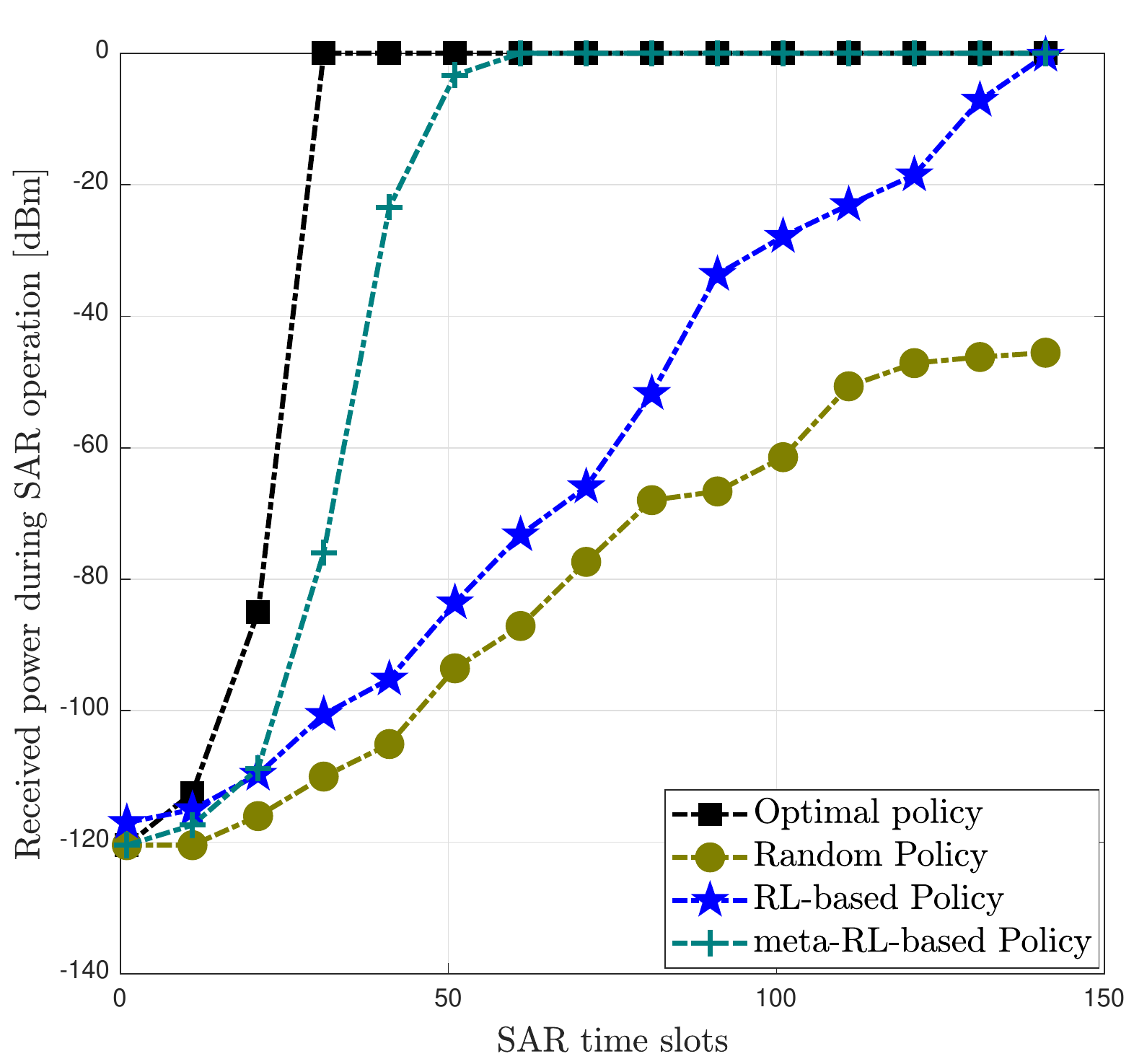}
 \caption{Received power at FL gateway v.s. rescue time slots in slotted canyon environment.}
   \label{Meta_Canyon_RSSI}
   \vspace{-5mm}
\end{figure}

In Fig.~\ref{Meta_Canyon_RSSI}, we show the received power at FL gateway versus rescue time slots in the slotted canyon environment. For the deep meta-RL policy, we have used the data of pervious experiences of different SAR operations in the plain environment, where the initial locations of UAV have been changed and the data of SAR has been saved in the memory. Then, this data is used as the $\mathcal{D}_{\text{Meta-train}}$ in Algorithm 2. From Fig.~\ref{Meta_Canyon_RSSI}, we observe that the received power at FL gateway under deep meta-RL policy receives to its maximum value at SAR time slot 50, while the deep RL policy at time slot 141 moves the FL gateway to the location with the maximum received power. Moreover, on average, the received power at the FL gateway under the deep meta-RL policy is $25\%$ more than deep RL policy.

\begin{figure}[t!]
  \centering
 \includegraphics[width=8cm]{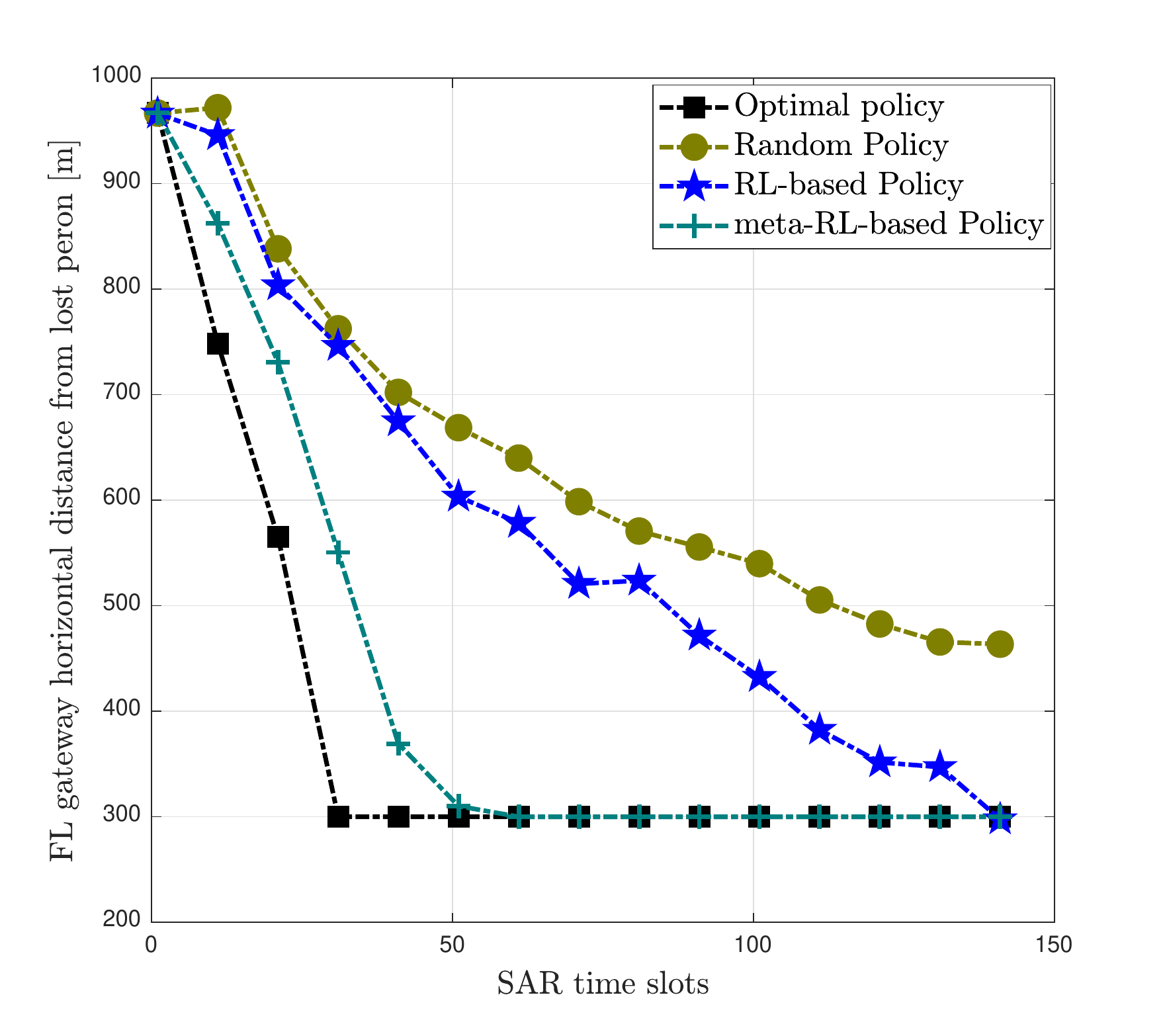}
 \caption{UAV horizontal distance from lost person in slotted canyon environment.}
   \label{Meta_Canyon_Dist}
   \vspace{-5mm}
\end{figure}

In Fig.~\ref{Meta_Canyon_Dist}, we show the FL gateway horizontal distance from lost person during SAR time slots in the slotted canyon environment. From Fig.~\ref{Meta_Canyon_Dist}, we observe that the deep meta-RL and deep RL based policies could finally find the lost person; however the greedy algorithm does not converge to the lost person location during SAR operation. As we can see in Fig.~\ref{Meta_Canyon_Dist}, the UAV hovers at $300m$ height over lost person after $31$ time slot under optimal policy. While deep RL and deep meta-RL-based policies find the lost person at time slots 51 and 141, respectively. Moreover, the FL gateway horizontal distance from lost person under deep meta-RL-based policy is on average $26\%$ closer than deep RL-based policy.

\begin{figure}[t!]
  \centering
 \includegraphics[width=9cm]{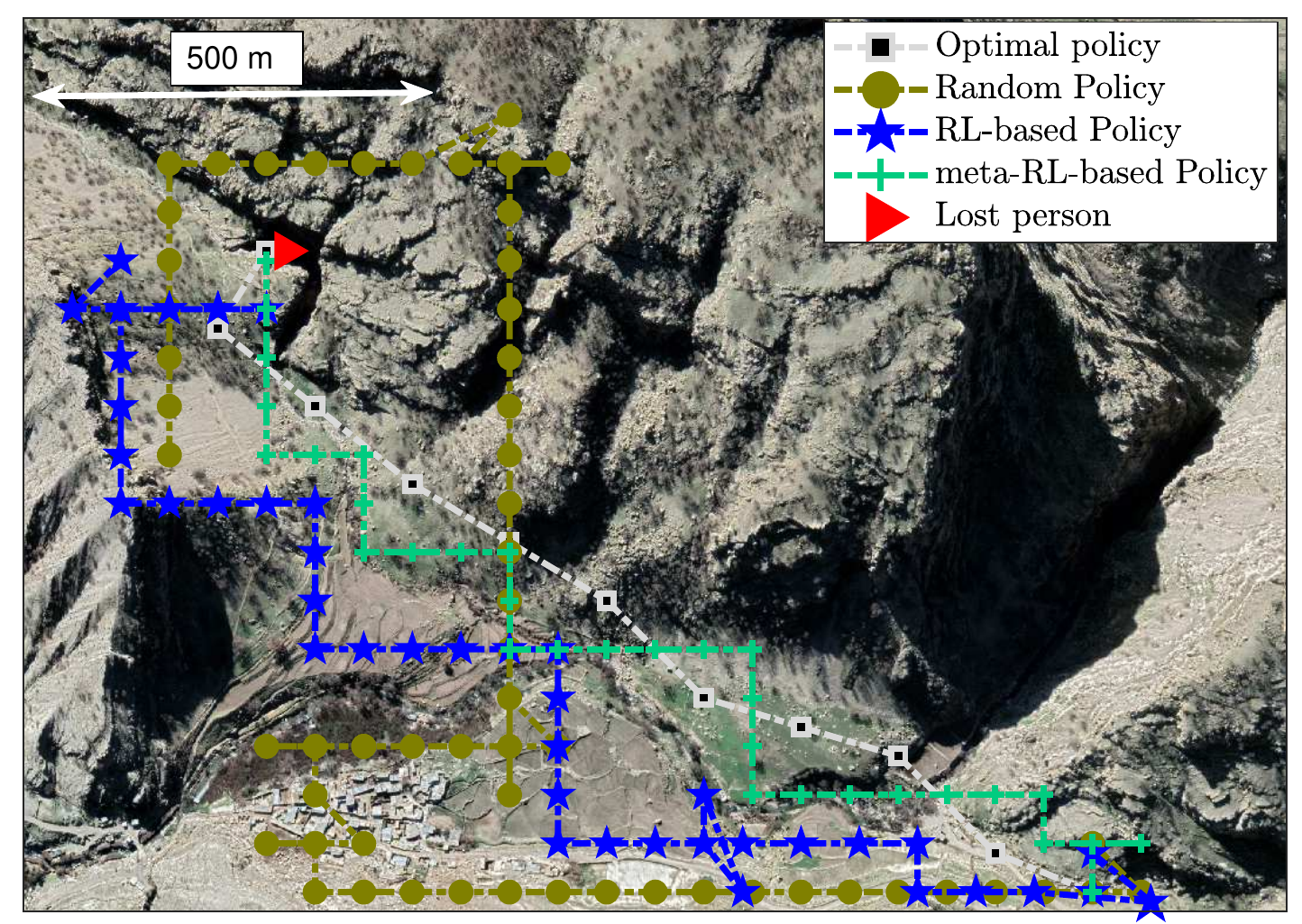}
 \caption{UAV trajectory during SAR operation at a slotted canyon. This satellite image is taken from
“Google Earth”. Lost person location is  $31^o35'18.24^{''}N$ and  $50^o 1'37.44^{''}E$.}
   \label{PracticalTest}
   \vspace{-5mm}
\end{figure}

In Fig.~\ref{PracticalTest}, we show the UAV trajectory during SAR operation at a slotted canyon. From Fig.~\ref{PracticalTest}, we observe that the UAV under the deep meta-RL and deep RL-based policies finally hovers over the lost person's location. The greedy algorithm moves the UAV towards lost person location while the UAV is not able to hover over exact lost person location during SAR operation time. As we can see in Fig.~\ref{PracticalTest}, the average distance between UAV trajectories under the deep meta-RL and deep RL based policies from the UAV trajectory under optimal policy are $619$ and $1930$ meter during the SAR operation time.
\section{Conclusion}\label{Sec:Conclusion}
In this paper, we have introduced a smart SAR system based on UAV-assisted LoRa network for highly remote areas such as mountain ranges. More prices, we have designed and implemented an artificial intelligence-empowered SAR operation framework for different unknown search environments. We have modeled the problem of the FL gateway control in the SAR system using the UAV-assisted LoRa network as a partially observable Markov decision process. Then, we have proposed a deep meta-RL-based policy to control the FL gateway trajectory during SAR operation. For initialization of our deep meta-RL-based policy, first, a deep RL-based policy determines the adaptive FL gateway trajectory in a fixed search environment including a fixed radio geometry. Then, as a general solution, our deep meta-RL framework is used for SAR in any new environment. Indeed, deep meta-RL-based policy integrates the prior FL gateway experience with information collected from the other search environments to rapidly adapt the SAR policy model for SAR operation in a new environment. We have experimentally implemented UAV-assisted LoRa network and then we have tested our proposed SAR system in two real different areas: a wide plain and a slotted canyon at Mongasht mountain ranges, Iran. Practical evaluation results show that if the deep meta-RL policy is applied instead of the deep RL one to control the UAV, the number of SAR time slots decreases from 141 to 50. Moreover, the average distance between UAV trajectories under deep meta-RL and deep RL based policies from the UAV trajectory under optimal policy are $619$ and $1930$ meter during the SAR operation time.
\bibliographystyle{IEEEtran}
\def\baselinestretch{0.9}
\bibliography{references}
\vspace{-1cm}
\end{document}